\documentclass[letterpaper,10pt]{article} 
\usepackage{opticameet3} %% use version 3 for proper copyright statement

\usepackage{amsmath,amssymb}
\usepackage[colorlinks=true,bookmarks=false,citecolor=blue,urlcolor=blue]{hyperref} %pdflatex
\newcommand*{\affmark}[1][*]{\textsuperscript{#1}}

\begin{document}

\title{Demonstration of Cooperative Transport Interface over Open Source 7.2 split RAN and Virtualised Open PON Network}

% \author{Author name(s)}
% \address{Author affiliation and full address}
% \email{e-mail address}
%%Uncomment the following line to override copyright year from the default current year.
%\copyrightyear{2024}
\vspace{-5mm}
\author{Merim Dzaferagic\affmark[1, *], Kevin O'Sullivan\affmark[2], Bruce Richardson\affmark[2], Brendan Ryan\affmark[2], Niall Power\affmark[2], Robin Giller\affmark[2], Marco Ruffini\affmark[1]}

\address{\affmark[1]CONNECT, School of Computer Science and Statistics, Trinity College Dublin, \affmark[2]Intel Corporation, Ireland}

\email{\{mdzafera, marco.ruffini\}@tcd.ie, \{kevin.o.sullivan, bruce.richardson, brendan.ryan, robin.giller\}@intel.com}

%\email{\authormark{*}fslyne@tcd.ie} %% email address is required

\vspace{-5mm}

\begin{abstract}
We demonstrate end-to-end 5G Open RAN over PON using off-the-shelf open networking hardware and open source RAN software. The implementation of the Cooperative Transport Interface provides timely synchronisation of PON and RAN schedulers. 
\end{abstract}
\vspace{-2mm}

\section{Overview}
\vspace{-2mm}
% An Overview section giving a comprehensive description of the demo's research theme and the technologies being demonstrated.

Open RAN is redefining network capabilities by providing an open and flexible framework to build the next generations of mobile networks. Open RAN is structured around three main components: the Central Unit (CU), Distributed Unit (DU), and Radio Unit (RU). The link between RU and DU/CU, typically referred to as fronthaul \cite{ngmn}, has traditionally relied on dedicated fiber networks. However, point-to-point fibre connectivity does not scale well economically with the principle of cell densification, which is a key milestone for capacity increase, through frequency reuse.
A more viable solution is instead the use of Passive Optical Networks (PON), already widely used for residential fibre broadband, which can offer ubiquitous fibre connectivity at a fraction of the cost, due to the highly shared nature of its infrastructure \cite{sharing}.

However, being a point to multi-point technology, PONs introduce additional latency upstream, as they run a Dynamic Bandwidth Allocation (DBA) protocol for capacity scheduling. Thus, in a typical fornthaul implementation, when packets are scheduled by the DU for the user equipment (UE), these are queued at the optical network unit (ONU) while upstream capacity is scheduled by the PON. This additional latency can break the RAN protocol when working in a functional split, such as the 7.2 configuration.
To tackle this, the concept of "Cooperative DBA" was first proposed in \cite{Tashiro2014}, aiming to develop an interface that enables coordination between RAN and PON schedulers. This was then standardised as Cooperative Transport Interface (CTI) \cite{itu2021}, enabling the DU in the RAN to inform the PON upstream scheduler a few ms in advance of upstream packets arrival from the UE. This allows the OLT to run the DBA algorithm in advance and schedule upstream grant allocations before packets arrive form the UE, reducing the packet queuing time at the ONU to a minimum.

%\textcolor{red}{Marco: the following sentence is not clear} The Medium Access Control/Physical layer (MAC/PHY) \cite{mountaser2017} functional split plays a pivotal role in this context, marking the interface between the MAC and PHY's Forward Error Correction (FEC).

A similar concept was recently demonstrated in \cite{CTI_nokia}, however this was over a proprietary RAN and proprietary PON from the same vendor. 
In this demonstration, we present for the first time, to the best of our knowledge, the implementation of the CTI interface in an open networking system for a 7.2 functional RAN split, based on: an open source DU/CU system, from the srsRAN project \cite{srsran}; an open source 5G core, from Open5GS \cite{5GS}; an Open RAN compliant RU, from vendor Benetel; open networking PON hardware, based on the Tibit transceivers, and open networking fully virtualised PON DBA implementation \cite{vDBA}.

This demonstration aims to showcase the ability to capture User Equipment (UE) grant requests from the DU and leverage this information to dynamically schedule PON upstream bursts via a software-driven virtual DBA. This approach enables adaptable bandwidth management, providing the necessary synchronization between PON and RAN schedulers to minimize latency and optimize network performance. The implementation leverages Intel technologies, including AF\_XDP and DPDK, for efficient, low-latency packet handling and high-speed data processing. %Additionally, we developed custom access mechanisms to interface with the Tibit Optical Line Terminal (OLT) architecture, using MongoDB to program grant schedules directly within the Tibit hardware.

\vspace{-2mm}

\section{Innovation}
\vspace{-2mm}
% An Innovation section describing the problem and the novelty of the demonstrated solution.
The key innovation of this demonstration is the implementation of a functioning CTI interface over a full end-to-end Open RAN over PON system, using off-the shelf open networking equipment: all hardware equipment (Tibit, RU and servers) are commercially available to any researcher, and the RAN protocol stack and 5G core are open source.

The demonstrated solution aligns the virtual Dynamic Bandwidth Allocation (DBA) mechanism in Passive Optical Networks (PON) with the 5G MAC scheduler using open networking frameworks. This approach addresses the rapid rise in demand for high-speed, low-latency data transmission in beyond-5G networks. Traditionally, the independent operation of PON and 5G NR systems limits their performance, causing resource allocation inefficiencies, data delays, and reduced Quality of Service (QoS). This integration is complex, as it requires coordination between different standards and technologies in both PON and 5G.

Our previous work \cite{ofc24} demonstrated the integration of the Tibit OLT DBA with srsRAN using a split 8 configuration, leveraging Tibit’s DBA cascading feature to coordinate scheduling across multiple OLTs. This approach underscored the benefits of sharing scheduling requests between the RAN and PON to reduce latency. However, due to the fixed bandwidth requirements of split 8—dependent on the RAN’s configured bandwidth—this setup could not fully exploit the potential for dynamic resource allocation in the fronthaul. 

Our current implementation moves to a split 7.2a configuration, which introduces dynamic bandwidth requirements in the fronthaul that vary based on the uplink traffic from user devices. This shift allows more adaptive resource allocation but introduces new challenges in synchronizing the Distributed Unit (DU) and Radio Unit (RU). Precise synchronization between the DU and RU is critical for 5G RAN operation. Given the high jitter in PON environments that affects the reliability of PTP, we implemented a GPS-based synchronization approach. Here, the RU directly syncs to GPS, while the DU synchronizes through a GPS-enabled PTP switch, ensuring accurate clock alignment across both units. Additionally, to account for increased fronthaul delay, timing adjustments are applied to the control and user-plane traffic at the DU.

In a non-cooperative scenario, a downstream 5G NR frame is sent from the CU/DU to the RU through the PON bearer network, which consists of an OLT-to-ONU fiber link, and then delivered to the UE over the radio network. This 5G NR frame includes an instruction for the UE to transmit upstream data. Typically, the PON does not prioritize this upstream 5G NR traffic. Therefore, a synchronization issue occurs on the upstream link when the ONU receives the 5G NR frame and places it in a TCONT buffer, along with other ONU applications, competing for an upstream transmission opportunity, known as the DBA opportunity time.  

\begin{figure}[h!]
   \centering
        \includegraphics[width=0.8\linewidth]{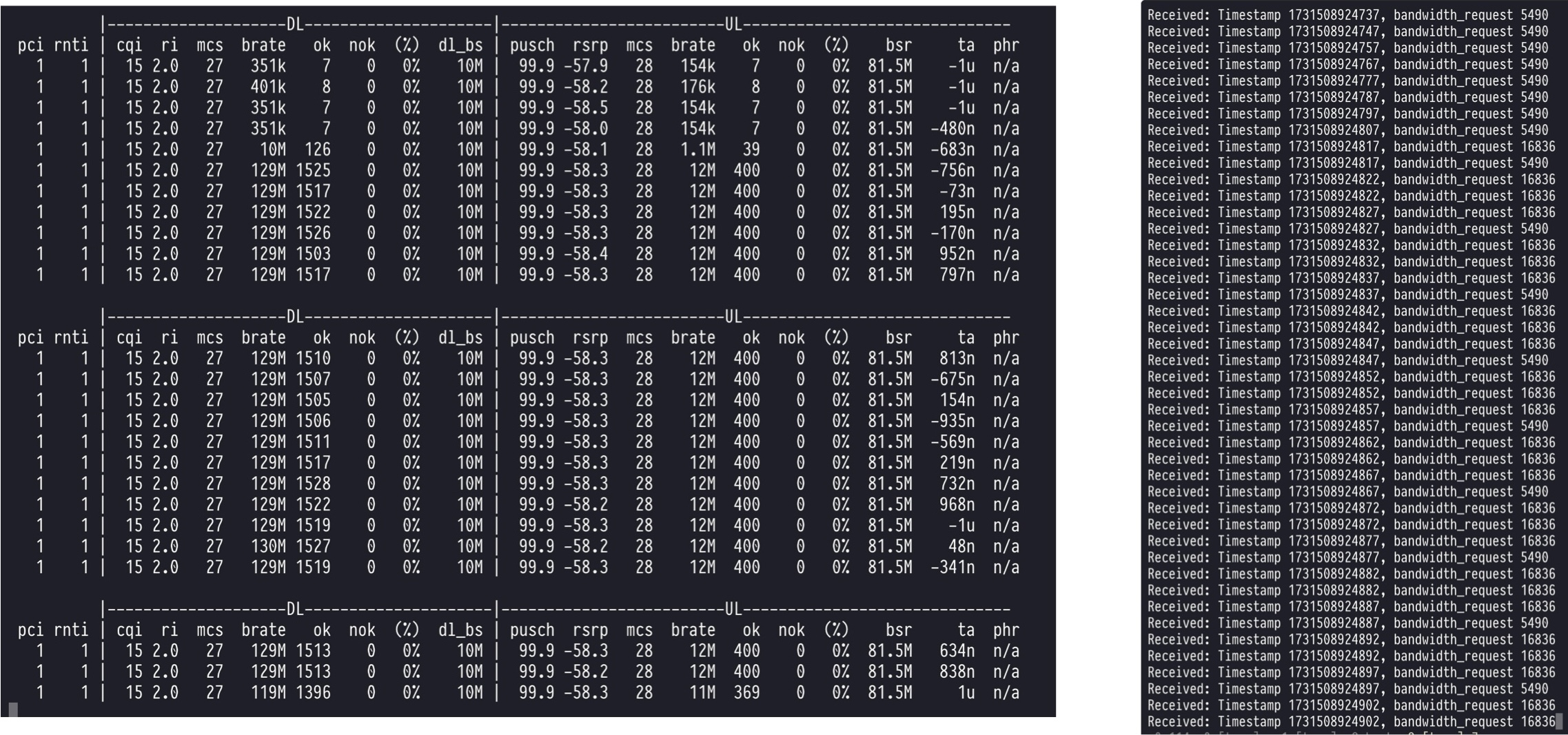}
        \vspace{-4mm}
    \caption{ (a) UE throughput and (b) CTI messages indicating the required bandwidth in the fronthaul. }
    \label{fig:Bandwidth_req}
\end{figure}

Our approach leverages UE upstream scheduling data available at the 5G DU to proactively align the PON DBA’s upstream scheduling. Figure \ref{fig:Bandwidth_req} illustrates this process: subfigure (a) displays the RAN scheduler terminal output showing user throughput, channel quality, UE bitrate, modulation and coding schemes and other RAN measurements for both uplink and downlink performance. Panel (b) shows the messages exchanged between the DU scheduler and PON over the CTI interface. Each message sent over the CTI interface corresponds to an information report from the DU to the UE, detailing the resources granted for a requested upstream transmission within a $1ms$ timeslot. These CTI messages initiate an upstream transmission from the associated ONU’s TCONT buffer connected to the O-RU. This notification, embedded in the downstream Bandwidth MAP (BWMap), enables synchronized and efficient scheduling of 5G NR traffic over the PON. As shown, PON bandwidth allocation dynamically adjusts to the real-time fronthaul requirements of the UE, optimizing resource usage across the network.
% On the other hand, we leverage UE upstream scheduling data available at the 5G DU to proactively align the upstream scheduling in the PON DBA. To illustrate our approach, Figure \ref{fig:Bandwidth_req} presents (a) terminal outputs showing the throughput experienced by the user, and (b) information on fronthaul requirements shared from the 5G MAC scheduler with the PON via the CTI interface. Each line in Figure~\ref{fig:Bandwidth_req} (a) represents the RAN performance measurements, including information about the channel quality, UE bitrate, modulation and coding scheme, etc in both, the uplink and downlink. Figure~\ref{fig:Bandwidth_req} (b) shows the messages exchanged between the DU scheduler and the PON over the CTI interface. Each line corresponds to an information report that the DU sends to the UE whenver the UE requests upstream resources. This infomration report contains information about the resources that are gratned to the user in a $1ms$ timeslot. The messages exchanged over the CTI interface initiate an upstream transmission from the associated ONU’s TCONT buffer connected to the O-RU. This notification is embedded in the Bandwidth MAP (BWMap), included as usual in the downstream frame, thus allowing synchronized and efficient upstream scheduling for 5G NR traffic over the PON. As illustrated in the figure, the bandwidth allocation within the PON dynamically adapts to match the actual fronthaul requirements from the UE.
\vspace{-4mm}
\vspace{-4mm}

% In this master/slave setup, one OLT serves as the “Master,” overseeing grant scheduling for the entire group. To handle real-time communication and optimize grant processing, we used ZeroMQ's (ZMQ) publish/subscribe communication pattern for low-latency exchange of information reports about the UE upstream scheduling decisions made by the DU. This ensures efficient coordination between the RAN and PON schedulers. We exposed the 5G MAC scheduler grants through the PHY Interface (FAPI), allowing MAC messages to be translated into messages for PON integration and optimized bandwidth allocation. A key element in managing diverse network traffic was the Intel X710 NIC’s traffic steering, which directed user-plane and DBA frames into separate queues, minimizing DBA buffering and scheduling delays. We further improved latency and throughput by using AF\_XDP (Address Family - eXpress Data Path) for high-performance packet processing that bypasses much of the Linux stack. 
\vspace{-2mm}
\section{OFC Relevance}
\vspace{-2mm}
% An OFC Relevance section explaining why the OFC audience and community should be interested in this demo.
Our demonstration aligns closely with key OFC concepts of open networks, virtualised access networks, and use of optical access technology to support viable densification of future 5G and 6G mobile networks. This demonstration emphasizes the convergence of optical access and mobile 5G technologies through open-source and open-networking frameworks, showcasing a flexible, software-driven approach to network management and orchestration. By employing RAN functional splits, specifically the 7.2 functional split over PON, we illustrate the evolution of RAN architectures to seamlessly integrate with optical networks. The use of the Tibit XGS-PON pluggable Optical Line Terminal (OLT) and Open RAN compliant commercial outdoor RU, further exemplifies innovation in open network hardware, offering a compact and effective solution for deploying PON infrastructure.

\vspace{-2mm}
\section{Demo content and Implementation}
\begin{figure}[h!]
   \centering
        \includegraphics[width=0.8\linewidth]{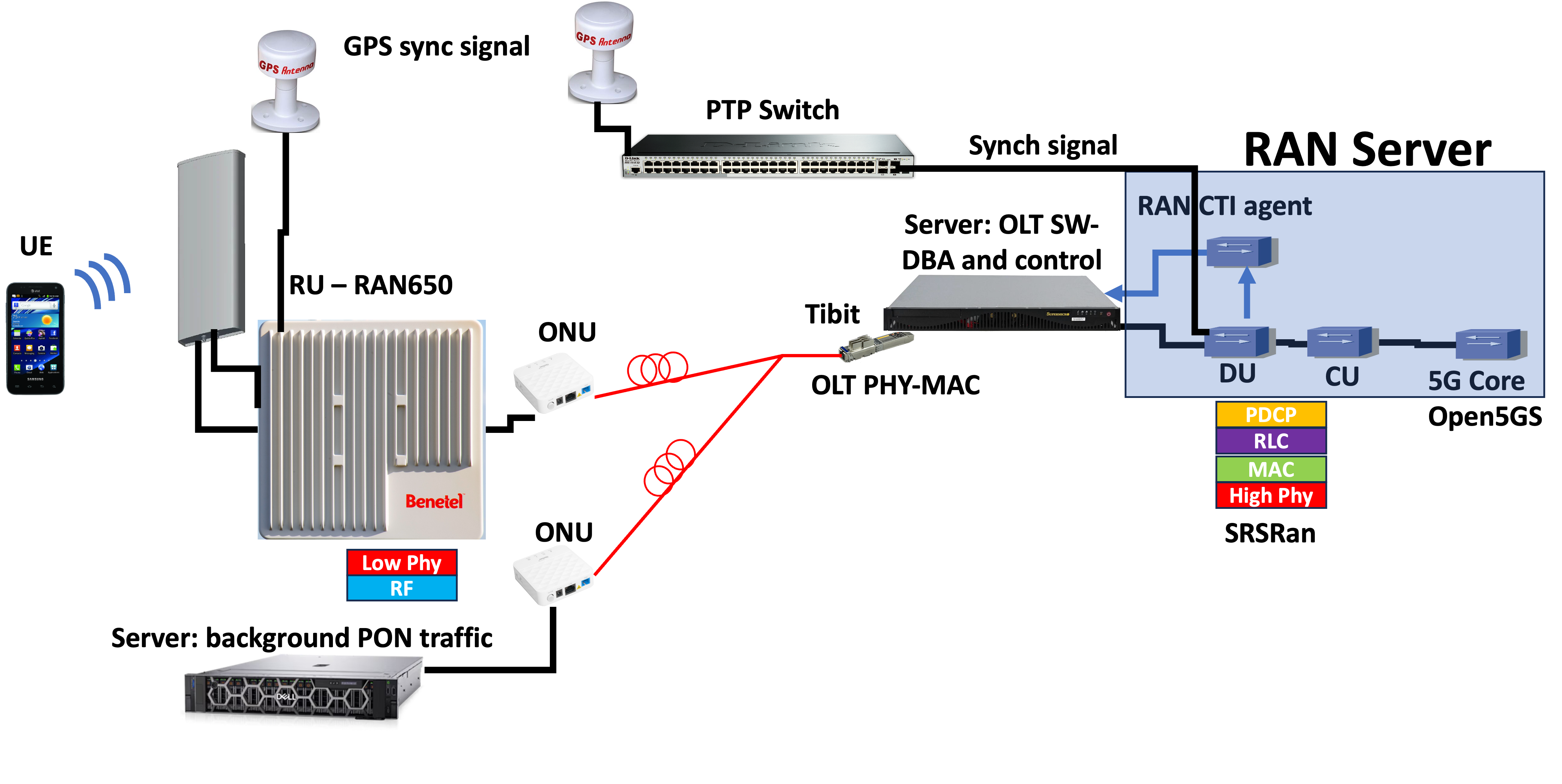}
        \vspace{-4mm}
    \caption{Demonstration Physical Architecture}
    \label{fig:Phys_arch}
\end{figure}

\vspace{-2mm}
%A Demo content \& implementation section clearly explaining:
%(i) the objectives and the configuration of the demo
%(ii) how the demonstration will be physically set up, including description of equipment to be used
%The demonstration will show the feasibility and interoperability of the CTI concept over PON, integrating PON DBA with 5G NR MAC scheduler. This will optimize network performance and highlight the importance of real-time data processing and communication in modern telecommunications networks.  
The objective of the demonstration is to showcase a fully functional 7.2 split Open RAN over PON, through open networking equipment. The demonstration will clearly show the benefit that activating the CTI interface provides, versus the use of a vanilla PON solution. 
The configuration includes a 5G new radio outdoor hardware RU working on the N77 band, connected by a PON, based on Tibit OLT, to an Intel-based server running the DU/CU and then 5G Core. The Tibit runs the PHY and MAC of the PON, while our virtual DBA implementation interacts with the DU scheduler, through our CTI implementation, to coordinate PON upstream scheduling requests. 
The physical setup that will be installed in the demo zone is shown in Figure \ref{fig:Phys_arch}.

Traffic will flow from a commercial smart phone running video streaming applications as well as dedicated traffic generator seamlessly towards the 5G core, which provides Internet connectivity.
The demonstration will also show telemetry data from the DU and PON, reporting metrics such as upstream latency and capacity utilisation, which will be made variable over time to emphasie the dynamic nature of the CTI interface.

\vspace{-4mm}
\vspace{-4mm}
%(iii) how the demo will be presented to the attendees
%We will demonstrate the enhanced network performance achieved by integrating the Cooperative Transport Interface (CTI) with PON Dynamic Bandwidth Allocation (PON DBA) and 5G NR MAC scheduler. 
%The physical setup, shown in Figure \ref{fig:Phys_arch}, includes a server simulating multiple User Equipment (UE) operations connected to the 5G NR MAC scheduler (part of the srsRAN project) via a high-speed Intel x710-DA2 Dual 10G Port NIC. Another server hosts the PON DBA, which is directly interfaced with the 5G NR MAC scheduler through CTI. The setup is interconnected using optical fiber links to ensure minimal latency and optimal data transfer. 
%Our logical setup will demonstrate how data flows from UE simulations to the 5G NR MAC scheduler, and how CTI facilitates effective communication between the 5G NR MAC scheduler and PON DBA. 
The demonstration will run in real-time, allowing attendees to observe the dynamic interaction between the 5G NR MAC scheduler and PON DBA, mediated by CTI. %This will provide a clear view of the differences in network performance, particularly in terms of latency and resource allocation efficiency, under scenarios with and without CTI integration. 
The session will begin with a slide presentation to set the stage for understanding the integration of CTI with PON DBA and 5G NR MAC scheduler. A live data display will offer a side-by-side comparison of network performance for optimized and non-optimized scenarios, with live application-level results and detailed performance measurements
\vspace{-2mm}
\section{Conclusion and Interaction}
\vspace{-2mm}
% (iv) how attendees might be able to interact with the demonstration (this last feature is essential to make the session engaging).
The presenters will first provide the audience with an overall concept of Cooperative Transport Interface, followed by a detailed overview of the algorithm and operational flow. Then the demo will run in real time, triggered by application on the smart phone, showing how the CTI will handle upstream capacity allocation for variable bit rate traffic. At the same time live metrics on latency and packet arrival will be collected and displayed, providing an immediate view of the system’s performance. Audience engagement will be encouraged throughout, with opportunities for Q\&A to delve deeper into the system’s functionality and offer hands-on insights. This format is designed to be both informative and interactive.

\section*{Acknowledgements}
\vspace{-2mm}
\small This work was supported by Science Foundation Ireland (SFI) under grants 18/RI/5721 and 13/RC/2077\_p2. This work was also supported by Horizon Europe SNS Grant 101139194.

\end{document}